\documentclass[usenatbib]{mn2e}
\usepackage[dvips]{graphicx}

\usepackage{amssymb}
\usepackage{natbib}

\begin{document}

  \title[Bars in cosmological simulations]{Bars in hydrodynamical
    cosmological simulations }

\author[Scannapieco \& Athanassoula]{Cecilia Scannapieco$^{1}$
and E. Athanassoula$^{2}$
\\
$^1$ Leibniz-Institute for Astrophysics Potsdam (AIP), An der Sternwarte 16, D-14482, Potsdam, Germany\\
$^2$ LAM, UMR6110, CNRS/Universit\'e de Provence,
38 rue Joliot Curie, 13388,  Marseille C\'edex 13, France}

   \maketitle

   \begin{abstract}
     
     We study the properties of two bars formed in fully cosmological
hydrodynamical simulations of the formation of Milky Way-mass galaxies. 
In one case, the bar formed in a system with  disc, bulge and 
halo components and is relatively strong  and long, as could be
expected for a system where the spheroid strongly influences the
evolution. The second bar is less strong, shorter, and 
formed in a galaxy with no significant bulge component. 
We study  the strength and length of the bars, the stellar density
profiles along and across the bars and  the velocity fields in the bar region.
We compare them with the results of dynamical (idealised) simulations
and with observations, and find, in general, a good agreement, although
we detect some important differences as well.
%On average, and taking into account the
%differences in mass and velocity distributions and in evolutionary
%history, we find good agreement. 
Our results show that more or less realistic bars
can form naturally in a $\Lambda$CDM cosmology, and open
up the possibility to study the bar formation process in a more
consistent way than previously done, since the host galaxies grow,
accrete matter and significantly evolve during the formation and
evolution of the bar.

   \end{abstract}


\section{Introduction}

Bars are present in roughly two thirds of all galactic discs 
\citep{Eskridge00,Barazza08} and can
drive the secular evolution of their host galaxies. 
They have thus been the subject of a number 
of studies based on dynamical simulations, i.e. simulations aimed
towards an understanding of the main relevant dynamical mechanisms
(e.g. \citealt{Combes90,Debattista2000,AM02, A03, MartinezValpuesta06}). 
These use idealised initial conditions, made specifically for the
question under study, and set aside all other effects, at least until an
understanding of the basic mechanism is achieved. The initial conditions
correspond to 
%Thus, in studies of
%bar formation and evolution, they use 
a fully developed
disc plus halo system which is as near equilibrium as
possible, so that the bar formation can be studied uninfluenced by other
instabilities. Furthermore, interactions with other galaxies,
major or minor, as well as inflow from the environment has been seldom
taken into account (see, however, 
%\citealt{Curir06,Curir07,Curir08}). 
\citealt{Curir06}), while the halo is usually
assumed to be initially spherical.
Moreover, the gas is generally
neglected or modelled without taking into account star formation,
feedback or cooling. 
%Although some studies did address the effects of
%one, or a couple of the above restrictions, no study so far has
%considered them all concurrently. 

In this letter, we take a different approach, i.e. we study
bar formation in the context of the
$\Lambda$CDM cosmology. In
this way, we include important external effects,
such as accretion and interactions and,  more important, the bars do not
wait for the galaxy to be fully formed  to start their own
formation. The disc and even the dark matter halo  keep growing while the bar
forms. Also, our galaxies do not have preset halo-to-disc mass ratios, velocity
dispersions, or other properties, and  all their properties directly result
 from
the simulations. This of course implies that we will not be able to
address the same questions as the dynamical simulations. For example,
 we can not  
examine how a given property of the host galaxy, e.g. the halo radial density
profile, will influence the formation and evolution of the bars.
%bar formation and evolution, since this would
%necessitate making a series of simulations varying this property only,
%while keeping all others fixed. 
On the other side,  we will include in our simulations
more physics that in any other single bar formation simulation. Our
main goal is to investigate if bars can naturally form in
$\Lambda$CDM, and compare their
% the basic properties of the
%bars in our simulations are realistic, and compare them with those of
%dynamic simulations. In a future paper, we will also compare
%our simulated bars with observational results.
properties with those of dynamic simulations and with
observations. 

The layout of the paper is as follows. In Section~\ref{sec:sims}
we describe the simulations, in Section~\ref{sec:results} we present
and discuss our results, and we conclude in Section~\ref{sec:conclusions}.

\begin{figure}
%\begin{center}
{\includegraphics[width=36mm]{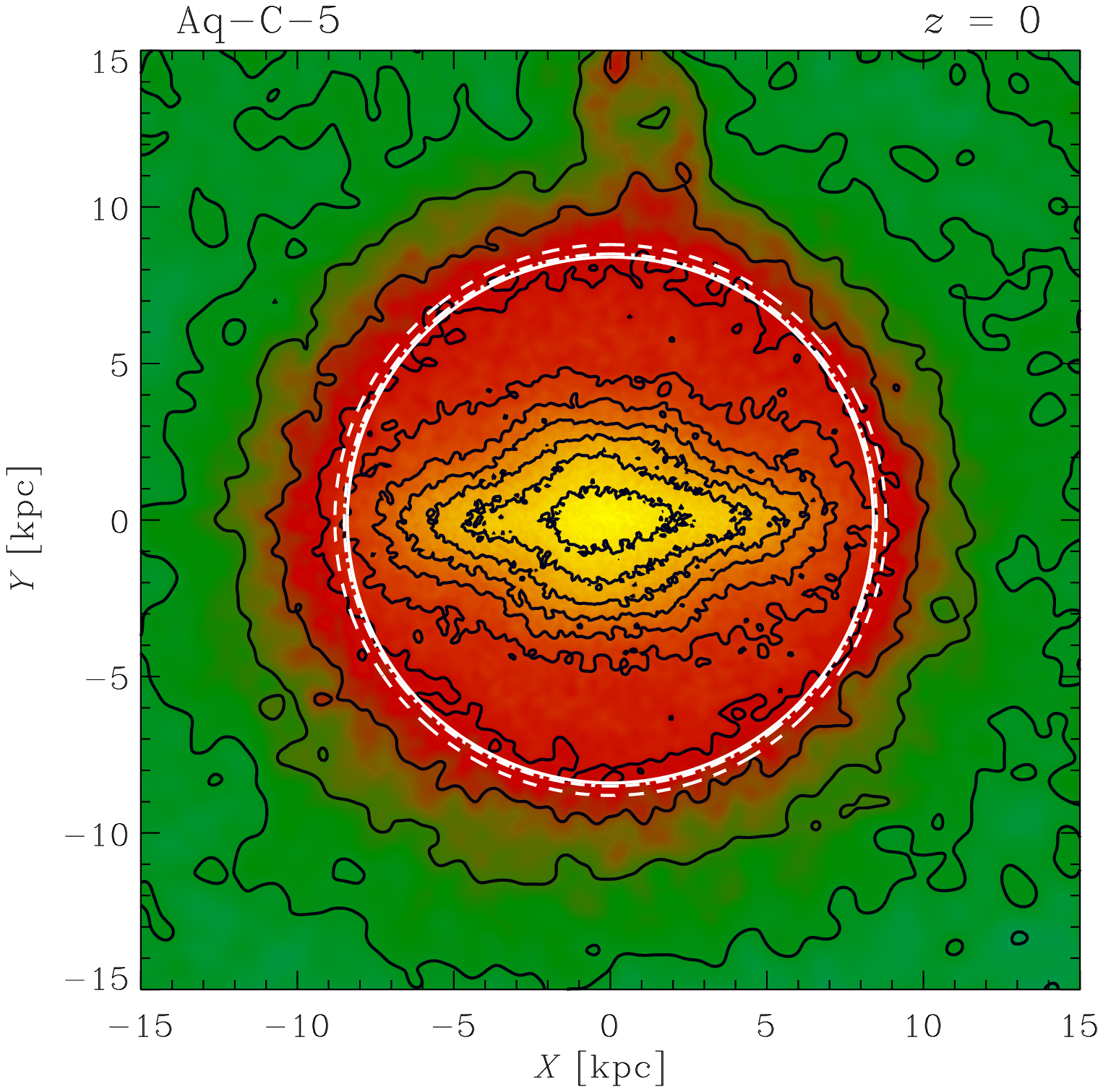}\includegraphics[width=36.mm]{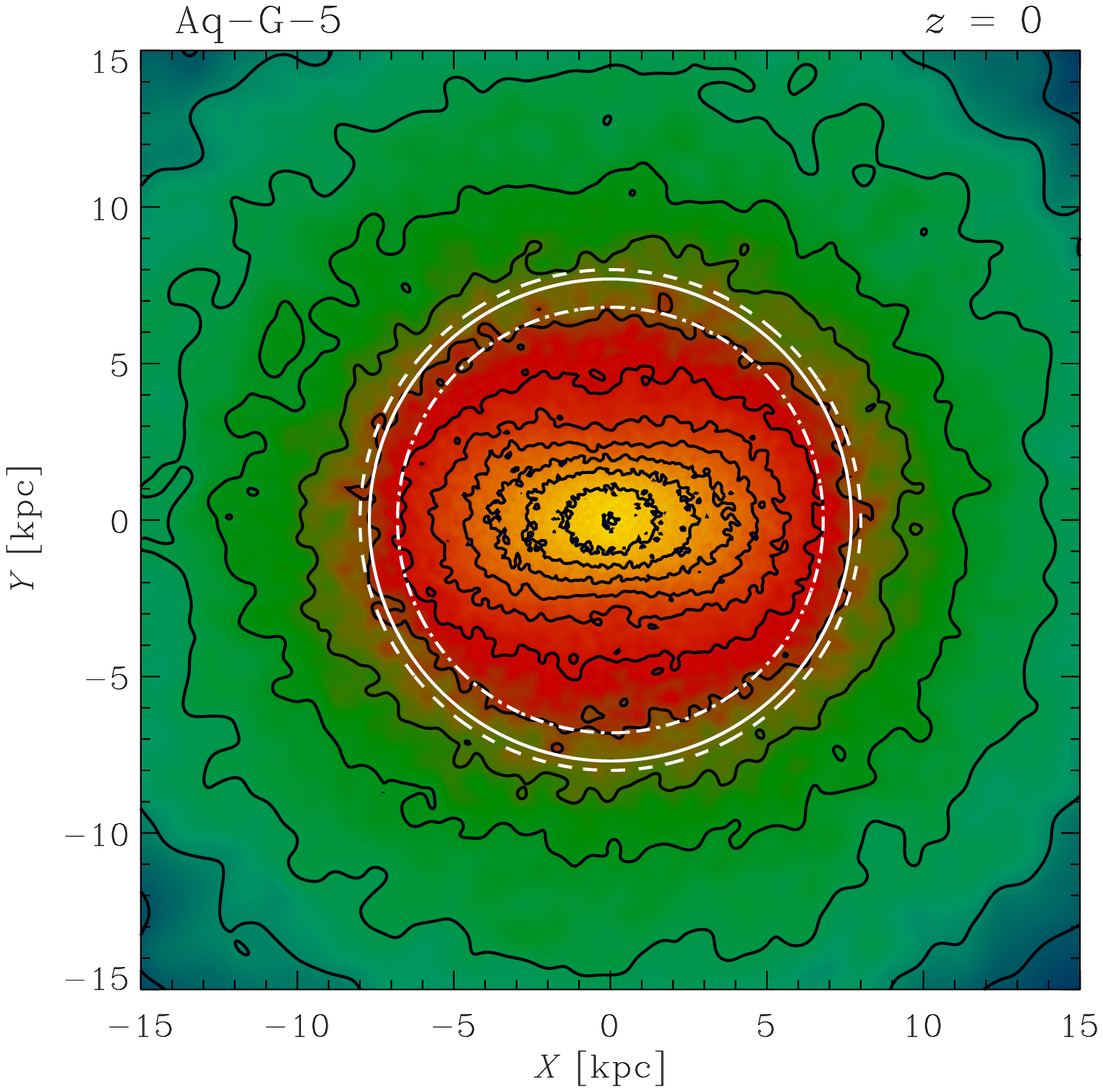}}

{\includegraphics[width=36mm]{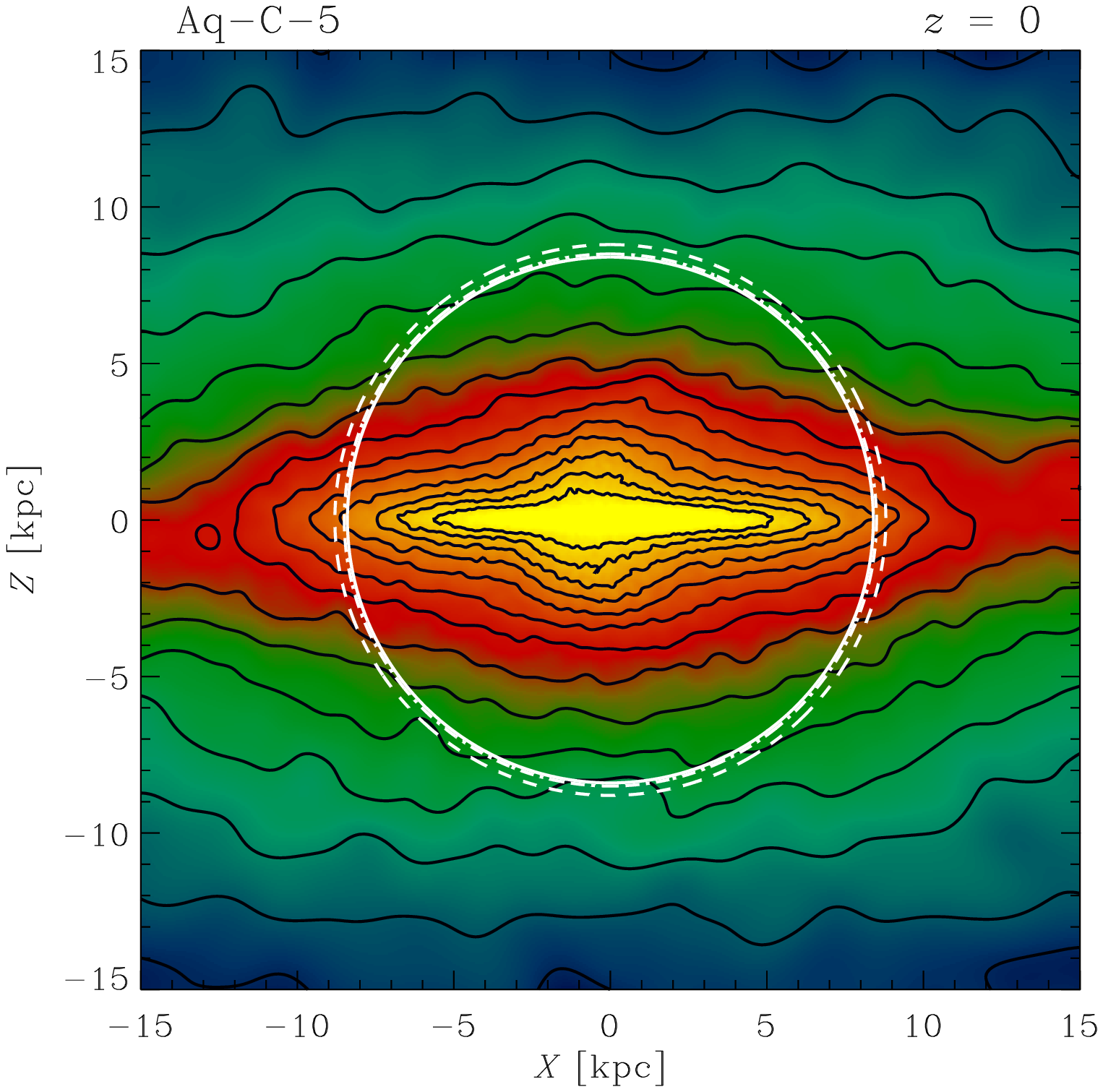}\includegraphics[width=36mm]{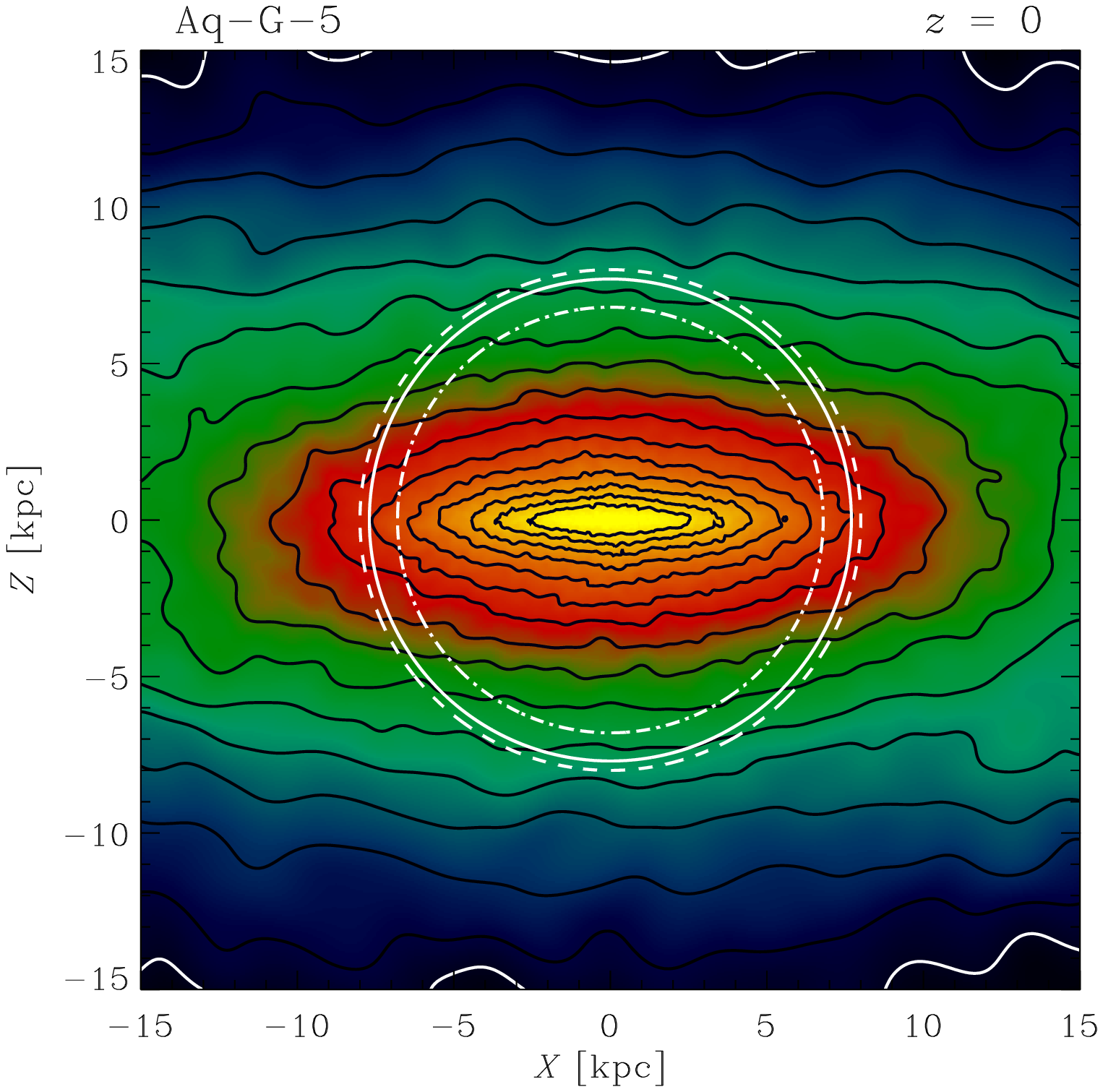}\includegraphics[width=36mm,angle=90]{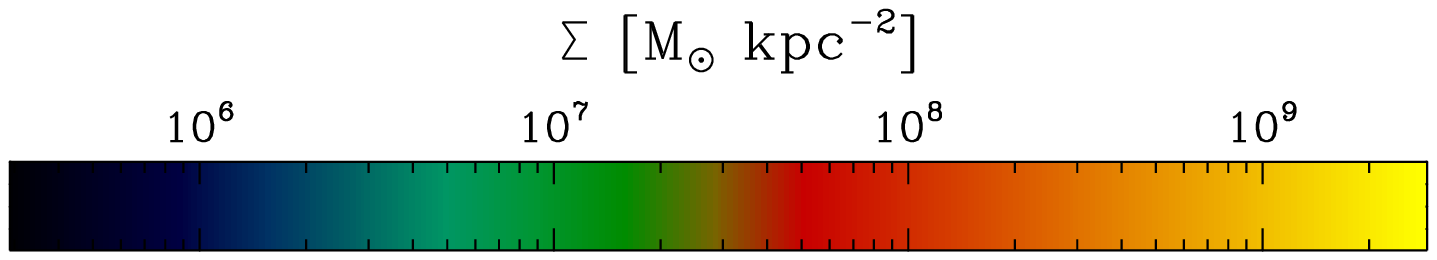}}

{\includegraphics[width=36mm]{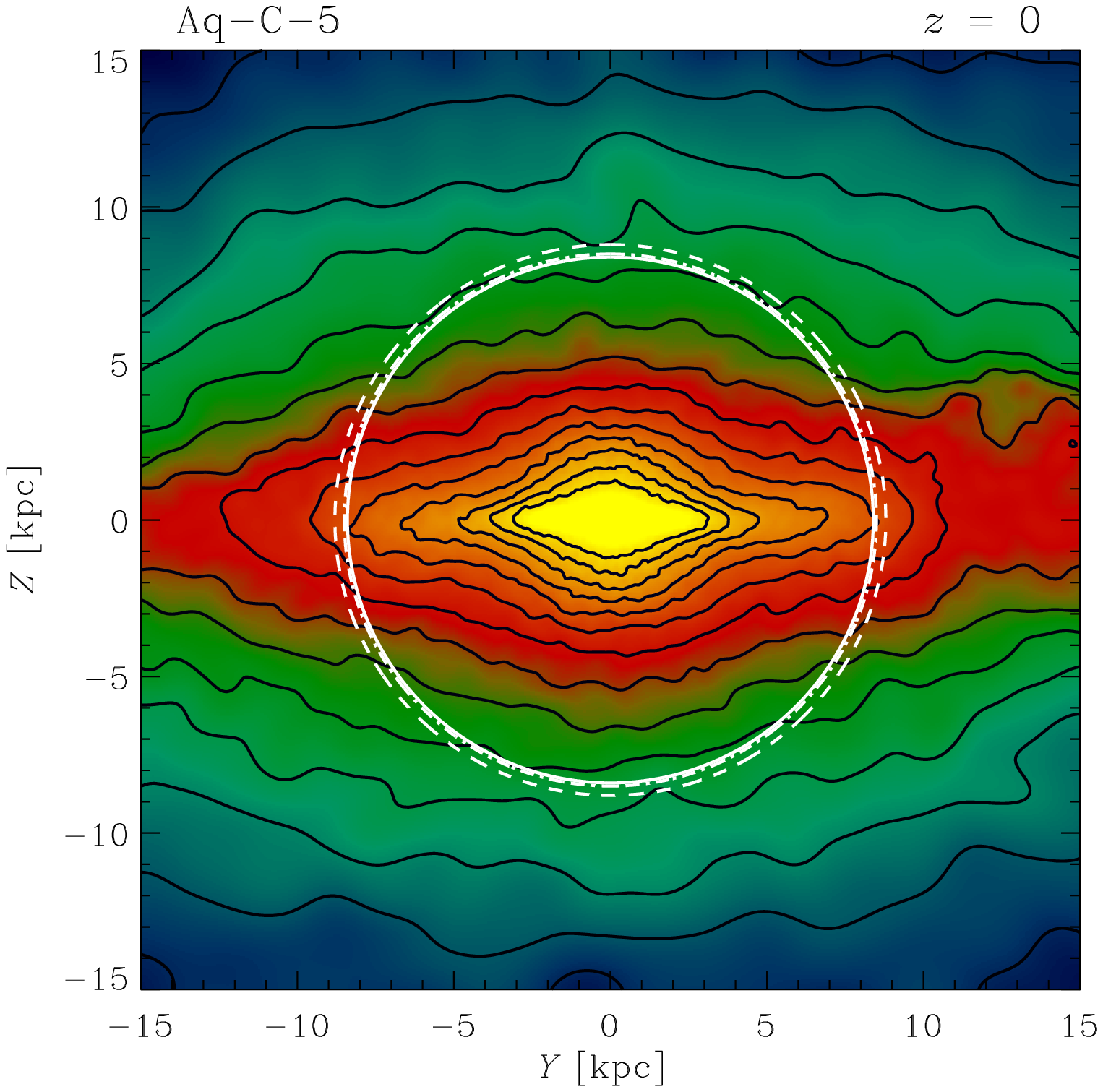}\includegraphics[width=36mm]{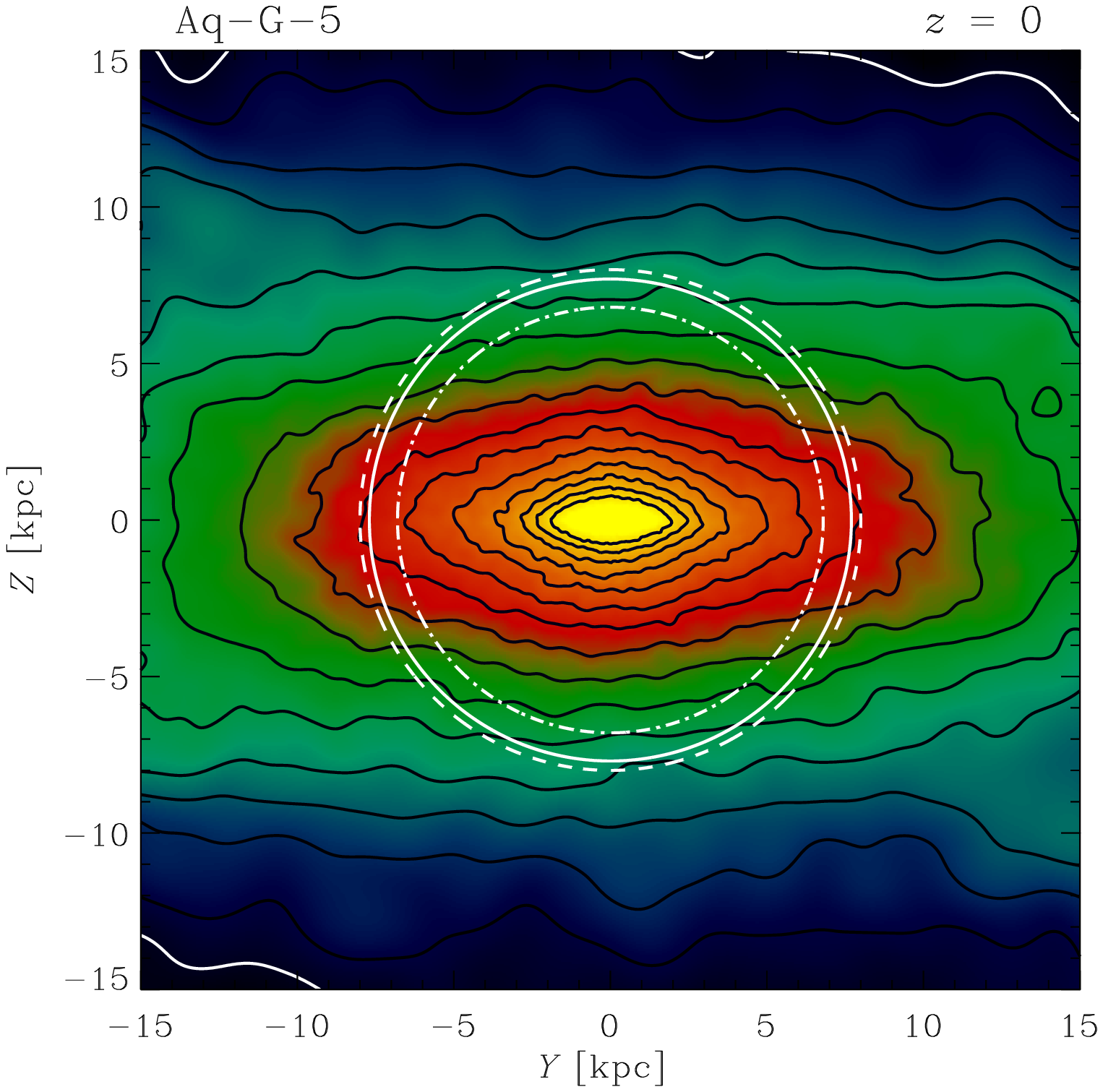}}

%\includegraphics[width=38mm,angle=0]{figures/ColorBar_isodensities.eps}
%\end{center}

\caption{Projected surface stellar density for Aq-C (left-hand panels)
  and Aq-G (right-hand panels). From top to bottom, the figures show the
  face-on ($XY$), side-on ($XZ$) and end-on ($YZ$)
 views.  The black lines correspond to isodensity contours equally-spaced in log$\Sigma$, and the white circles indicate the bar length obtained with three different
methods (see Section~\ref{sec:length}). }
\label{fig:isodensities_z0}
\end{figure}

\vspace{-0.2cm}
\section{Simulations and analysis}
\label{sec:sims}

We study the present-day properties of bars formed
in two hydrodynamical simulations of galaxy formation in a $\Lambda$CDM
universe. The simulations correspond to galaxies that, at redshift $z=0$,
are similar in
mass to the Milky Way and are mildly isolated. They are a
sub-sample of the eight simulations extensively described  in 
\citet[hereafter S09, S10 and S11, respectively]{S09,S10,S11}.
%\citet[][hereafter S09]{S09}, 
%and extensively
%described  in S09, \citet[][hereafter S10]{S10} 
%and \citet[][hereafter S11]{S11}.
We analyse here the galaxies named Aq-C (Aq-C-5 in S09, $M_{200} = 1.6\times 10^{12}$M$_\odot$)
 and Aq-G (Aq-G-5 in S09, $M_{200} = 6.8\times 10^{11}$M$_\odot$),
 where bars are clearly present at $z=0$. 
Two other galaxies, Aq-A and Aq-E,
have bars at $z=0$ (S10); however, we defer their analysis to a separate
work since the evolution of these systems is far more complicated. Aq-A has
two stellar misaligned discs at $z=0$, while Aq-E has a rotating
bulge and a recent interaction with a satellite galaxy which strongly
disturbs the system (S11).

We assume a 
$\Lambda$CDM cosmology with 
$\Omega_\Lambda=0.75$, $\Omega_{\rm m}=0.25$, $\Omega_{\rm b}=0.04$,
$\sigma_8=0.9$ and $H_0=73$ km s$^{-1}$ Mpc$^{-1}$. 
The mass resolution is $\sim 10^6$M$_\odot$ for
dark matter and $\sim 2\times 10^5$M$_\odot$ for baryons (S09).
The gravitational softening  is $1.4$ kpc, it is fixed in comoving coordinates 
and is the same for gas, stars and dark matter particles.

The simulations were run with  the Tree-PM SPH
code {\small GADGET-3} \citep{Springel08}, with the additional 
modules of \citet{S05, S06}.  It
includes star formation, chemical enrichment and (type II and
Ia) supernova feedback, metal-dependent cooling, 
an explicit multiphase model for the gas component, and the effects
of a UV background. 
We refer the reader to S09 and S11
for full details on the initial conditions, 
simulation code and resolution effects.

Our analysis methods largely follow those in
\citet[][hereafter AM02]{AM02}, to
easily compare our results to   dynamic simulations
and  to observations. There are, however,
two  differences. First, unlike in dynamic simulations
where it is possible to analyse discs separately from bulges
(all through this paper,
 by bulge we mean classical bulge, 
  \citealt{KK2004}), here they
constitute together the stellar component and any attempt to
distinguish between them is approximate. 
%Although this
%prevents us from making some of the useful comparisons with previous
%simulations, it allows direct comparisons with observations,
%since bulge/disc distinction is not straightforward there either. 
The second difference is that
in dynamic simulations the softening is
about one tenth
of that used here. Consequently, our results will be more
smoothed out than those of dynamical simulations.
% and this should somewhat
%influence several quantities.

We will also follow the terminology introduced in AM02, where MH
denotes simulations with strong bars, in which the near-resonant
material in the bar region emits a considerable amount of
angular momentum, which is absorbed mainly by the near-resonant material in
the halo. Simulations in which considerably less angular momentum has
been redistributed, and which therefore
have less strong bars (\citealt{A03}, hereafter A03), are called MD.

\vspace{-0.54cm}\section{Results}\label{sec:results}

\subsection{Morphology}
\label{subsec:Morphology}

Fig.~\ref{fig:isodensities_z0} shows maps of projected surface stellar 
mass density for Aq-C and Aq-G.
The projection is such that  $XY$ is the
disc's equatorial plane, with angular momentum
in the positive $Z$ direction and with the bar major, intermediate and
minor axes lying along the $X$, $Y$ and $Z$ axes, respectively.

Aq-C has a significant well-defined disc, a
bulge-like component and a bar. 
The bar looks very strong, very thin and rectangular-like.
Seen face-on, the bulge has a rather peculiar shape. In contrast, Aq-G
does not have a significant bulge (S10),
%, in good agreement with the photometric decomposition of S10,
but has an important bar and a clearly identifiable disc.
Its bar is very symmetric and, viewed face-on, much fatter
than that of Aq-C.

We find some differences between our simulated galaxies
and both real galaxies and
galaxies in dynamic simulations, in particular in the
face-on views. We find that Aq-C has a bizarre shaped bulge, unlike
what is usually observed. Note however that, in a $\Lambda$CDM cosmology,
the shape of galaxies can change very rapidly particularly
during and/or after interactions with satellites and mergers. 
In the case of Aq-G, the disagreement is quantitative, rather
than qualitative. 
Rectangular-shaped bars have been many times
observed, but always in strong bars \citep{A90,AM02, Gadotti11}.
On the other hand, the bar in Aq-G is fat, more like
an oval (i.e. weak),
but still has a clear rectangular outline. 
Aq-G then looks like a hybrid between strong and weak bars.

\begin{figure}
\begin{center}
{\includegraphics[width=38mm]{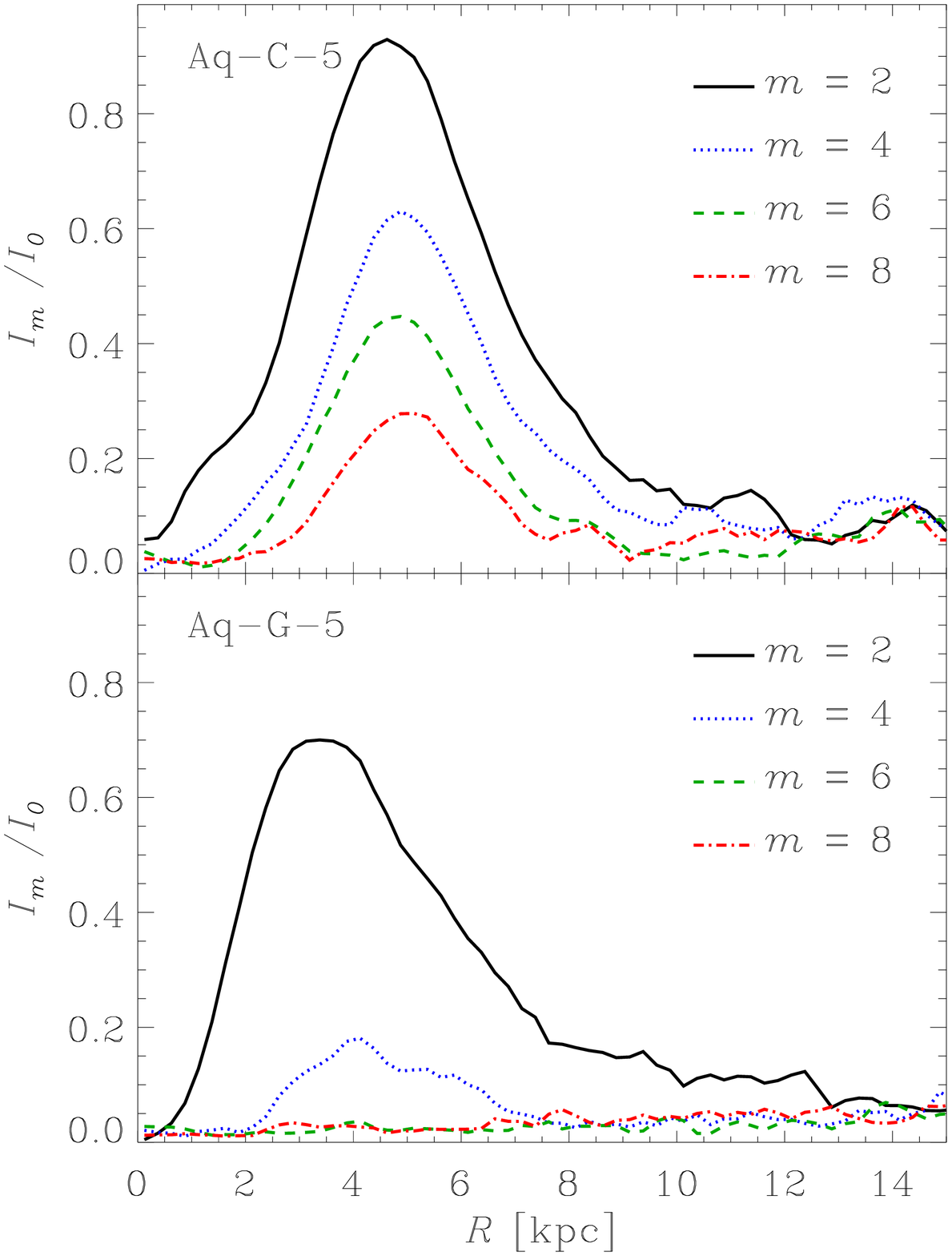}\includegraphics[width=38mm]{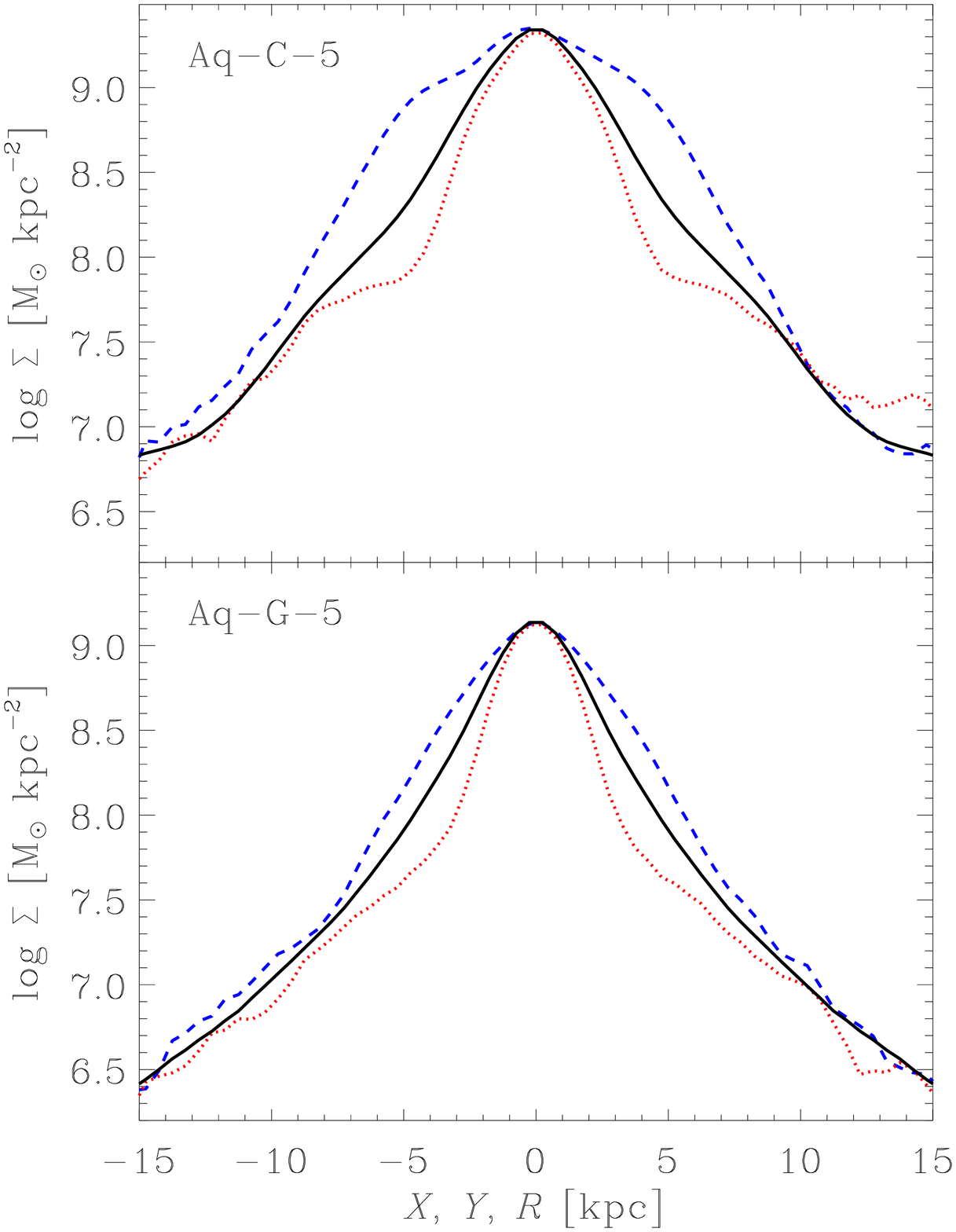}}

\caption{{\it Left:} Relative Fourier amplitudes of the even components up to $m=8$. 
{\it Right:} Projected mass density profiles along the bar major axis ($X$, dashed lines), the bar 
minor axis ($Y$, dotted lines), and  azimuthally averaged 
(solid lines), when the galaxy is seen face-on. } 
\label{fig:all}
\end{center}
\end{figure}

\subsection{Bar strength}
\label{subsec:bar-strength}

In order to quantify the bar strength
we Fourier analysed the projected face-on mass density ($\Sigma$), i.e.:
\begin{equation}
\Sigma(R,\theta) = {A_0(R)\over{2}} + \sum_m [A_m(R) {\rm cos} (m \theta) + B_m(R) {\rm sin} (m \theta) ]
\end{equation}
with $\theta$ being the azimuthal angle
and $R$  the cylindrical radius.
In practice, we calculate $A_m(R)$ and $B_m(R)$ as:
\begin{equation}\label{f1}
A_m(R) = \sum _{i}~m_{i}~{\rm \cos} (m\theta_i), ~ m\ge 0
\end{equation}
\begin{equation}\label{f2}
B_m(R) = \sum _{i}~m_{i}~{\rm \sin} (m\theta_i), ~ m>0
\end{equation}
where 
$m_i$ is the mass of the stellar particle $i$ and the summation
is over all particles in an annulus around radius $R$. We define  
the amplitudes of the Fourier components as $I_m=\sqrt{A_m^2+B_m^2}$ ($m>0$)
and $I_0=A_0/2$,  and use their ratio as a measure of the bar strength (note 
that several dynamical simulation studies and observations use the
ratios $I_m/A_0$ instead of $I_m/I_0$.). We show $I_m/I_0$  for Aq-C
and Aq-G (up to $m=8$)  
in Fig.~\ref{fig:all} (left-hand panels).

Aq-C, the strongest of our two bars, has a large $m=2$ component.
Its maximum, $(I_2/I_0)^{\rm max}=0.92$, occurs at
$R_{\rm max} = 4.5$ kpc and is similar to those found in dynamical 
simulations with gas but no star formation
(\citealt{VillaVargas2010} and references 
therein). Compared to dynamical simulations with star
formation, feedback and cooling, Aq-C corresponds roughly to 
simulations with most baryons initially in a gaseous component and
reaching about 6 to 10\% gas in the disc at times comparable to $z =
0$ (Athanassoula et al., in prep.).
% and is also in the range of values obtained for observed galaxies. 
The shape of the radial profile is  more reminiscent of that found
in gas-rich dynamical simulations. Some of the differences could,
nevertheless, be due to the 
fact that we are including in the Fourier analysis all stars,
i.e. from the disc, the bar and the bulge. This will necessarily lower 
the $m$ components, though, in this case at least, not all $m$ in the same
way since the bulge has a clear rectangularity. Also, $m$ = 4 aside,
this lowering must be stronger in the inner parts, where the bulge
contribution is larger.   
 
The higher order even moments  ($m=4,6,8$) have maxima of  $I_4/I_0=0.64$, 
$I_6/I_0=0.44$ and $I_8/I_0=0.28$, 
%i.e. are also very high, 
in very good agreement with MH-type simulations. 
These maxima occur at
radii roughly equal to that of the $m=2$, while in MH-type 
dynamical simulations they occur considerably further out. This
also could be due to the strong rectangularity of the bulge
 (Fig.~\ref{fig:isodensities_z0}). 

The bar of Aq-G is less strong, with $(I_2/I_0)^{\rm max}=0.70$ 
that occurs at $R_{\rm max}=3.4$ kpc. 
The $m=4,6,8$ moments are very small ($<0.18$) at
all radii, in good agreement with the more elliptical-like bar outline. 
In fact, the $m>4$  components are within
the noise. Thus they are in many ways similar to those from MD
simulations, where less angular momentum has been
redistributed within the galaxy than in MH-types (AM02, A03). 
The fact that $R_{\rm max}$ is shorter in Aq-G than in Aq-C
indicates that the bar is shorter, as already inferred
from Fig.~\ref{fig:isodensities_z0}. Note, 
however, that the two galaxies have neither the same mass, nor the
same extent, as seen in Fig.~\ref{fig:isodensities_z0} (see also
Section~\ref{sec:length}).

Another important  difference between the two bars is the shape
of the $I_2/I_0$ amplitude for $R<R_{\rm max}$ and $R>R_{\rm max}$. 
In Aq-C, the declines on the two sides of $R_{\rm max}$ are similar, whereas
in Aq-G the decline is clearly steeper inwards.
In dynamical simulations, the norm is an asymmetrical decline on the
two sides of the maximum (but
see simulations MH1 in \citealt{ALD2005}, which has
a strong central mass concentration). 
%This is
%also the case in most observed galaxies, with, again, some notable
%exceptions such as NGC 1387, 1440 or 3941 \citep{Buta2006}.

Finally, the odd Fourier components ($m=1,3,5,7$) are, both
for Aq-C and for Aq-G, very small ($<0.1$) at all radii
indicating that both bars are quite symmetric.

Our results are in relatively good agreement with observations, 
that find a great variety of $I_2/I_0$
profiles and of maximum values, typically between  
$0.4$ and $0.8$ (\citealt{Ohta90,Aguerri2001,Aguerri2003,Buta2006}).
In Aq-C,  $I_2/I_0^{\rm max}=0.92$ is very high, but values
as high as $0.9$ have
also been observed \citep{Buta2006}.

\subsection{Density profiles along major and minor axes}

Fig.~\ref{fig:all} (right panels) shows the stellar density
profiles viewed face-on, along the major ($X$) and minor ($Y$) axes
of the bar, as well as azimuthally averaged
(note that we include here all stars
in the simulations, i.e. bulge, disc and bar).
%, so that our results are
%directly comparable to observations.

The profiles of the two simulations differ significantly between them,
while having a number of similarities with those of    
dynamic simulations. The profile of Aq-G along the bar major axis shows a clear 
near-exponential drop; a behaviour similar to that of MD models,
and to the `exponential' bars often observed in late-type galaxies \citep{EE1985}.  
In contrast, Aq-C has a less typical profile. It shows a 
flattish section, but of short extent, arguing that it is 
linked to the bulge rather than the bar, contrary to the flattish parts 
of MH bars (Fig.~5 in AM02) 
and of 'flat' bars in observed early-type galaxies  \citep{EE1985}. 
Furthermore, and 
contrary to MH-type bars in dynamic simulations, there is no abrupt fall
of the density profile at the position of the bar
end,  presumably due to the fact that the softening in the
cosmological simulations is much larger and so washes out the
relatively steep drop at the end of the bar.

The density profiles along the minor axis of the two galaxies also
show significant differences. In  Aq-C, the profile shows 
a clear change in slope at the position where the bar ends. This
behaviour is also found in MH-type bars and
means that
the bar is sufficiently strong to clear out the region around the
Lagrangian points, as one sees in many
observed strong barred galaxies. Aq-G also shows a change in the slope
of the density profile along the bar
minor axis; but much less pronounced than Aq-C.
Fig.\ref{fig:all} also shows that the difference between the cuts along
the major and minor axes in the bar region is considerable larger in
Aq-C than in Aq-G, confirming that the bar is much stronger
in the former case.

\begin{figure}
\begin{center}
{\includegraphics[width=40mm]{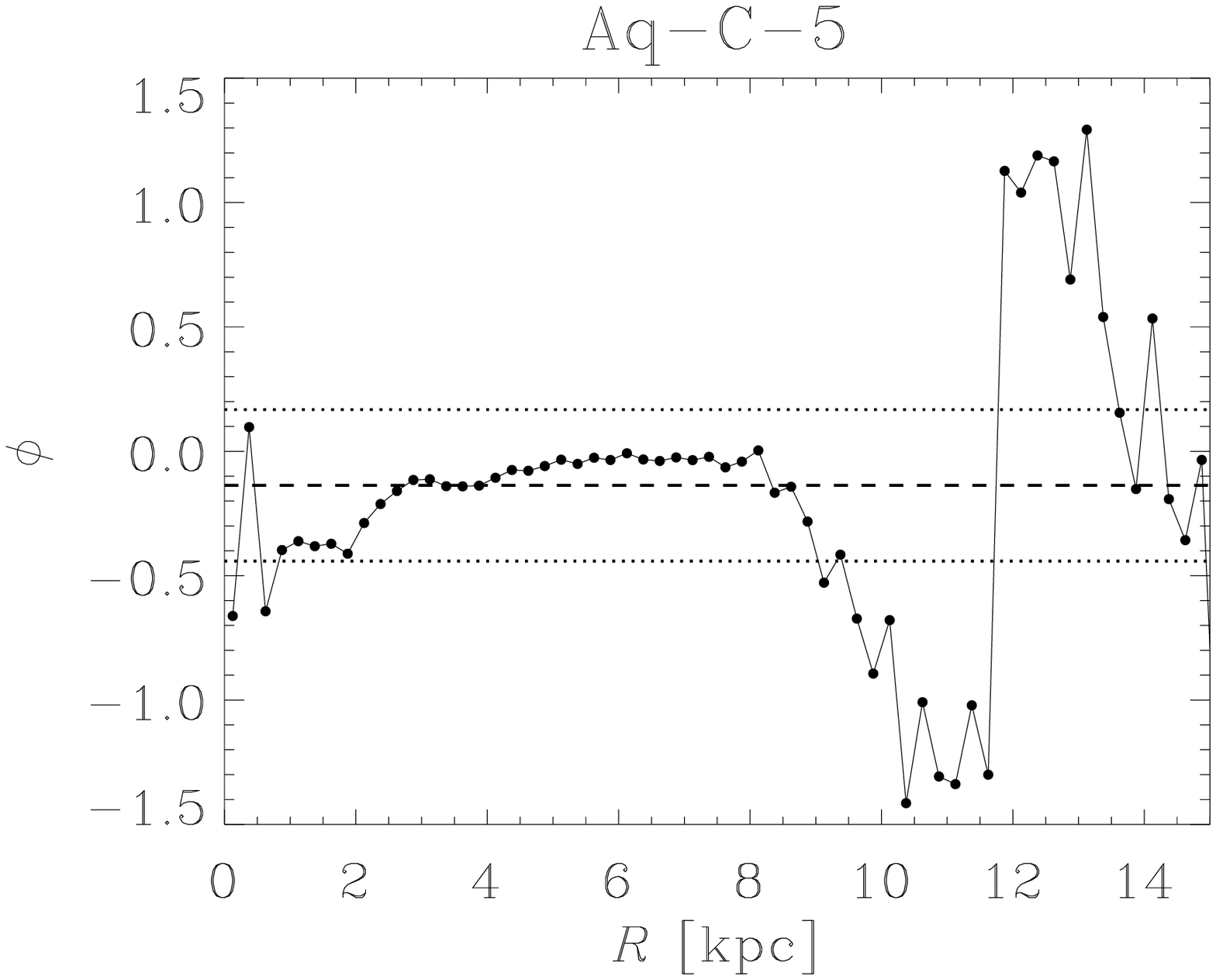}\hspace{-0.2cm}\includegraphics[width=40mm]{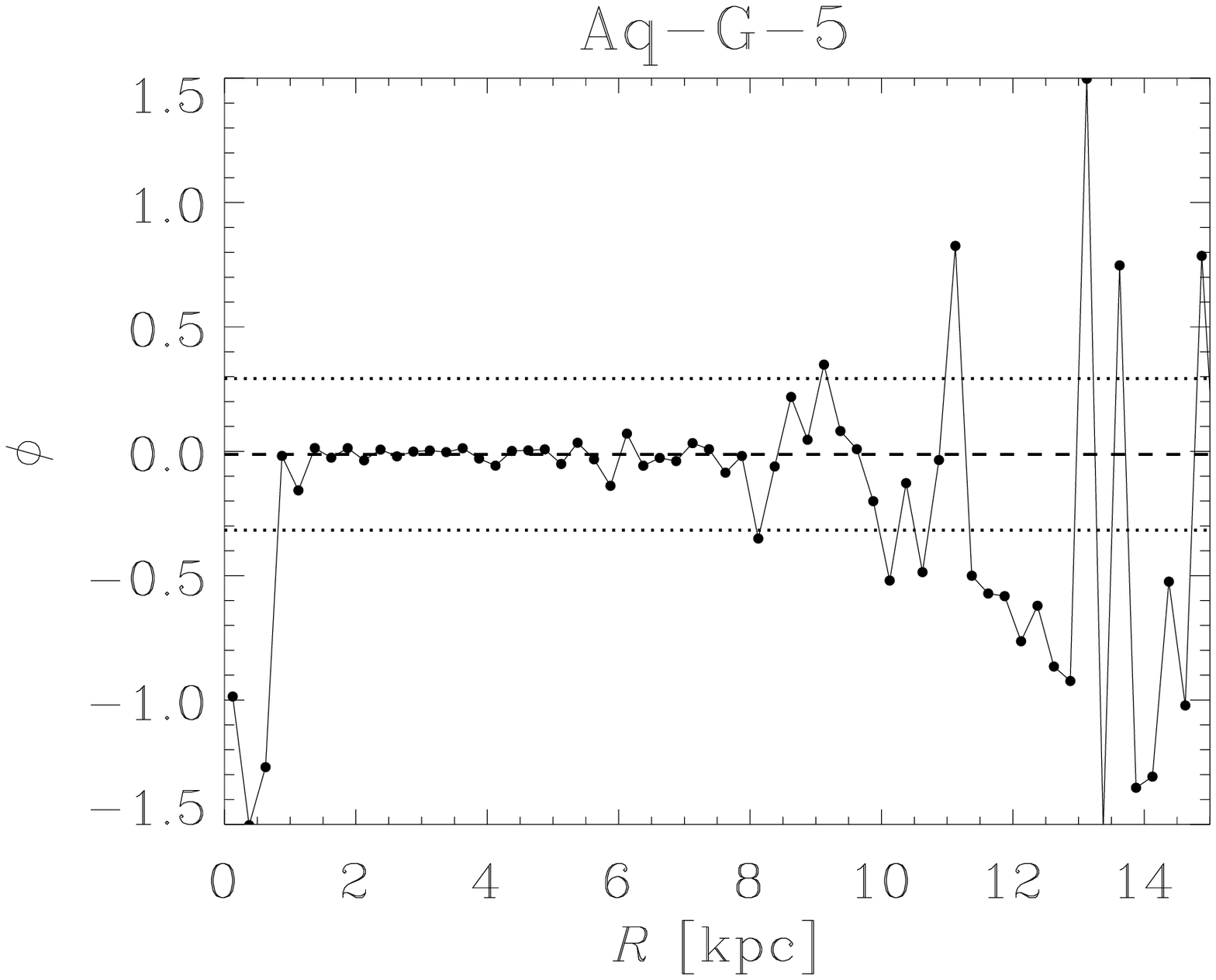}}
\caption{ Phase of the stellar distribution as a function of radius. The dashed lines 
indicate the phase of the bar ($\phi_{\rm b}$), and the dotted lines are at $\phi_{\rm b}\pm$ arcsin$(0.3)$.}
\label{fig:phase}
\end{center}
\end{figure}

\subsection{Bar length}\label{sec:length}

AM02 presented a number of ways to measure bar length and
discussed their advantages and disadvantages. Whenever possible or
useful, these methods were extended to observed bars \citep{Gadotti07}. As not
all of these methods are well adapted to cosmological simulations, we
will here use three of them, of which one after a small modification. 

The first method 
({\it iii} in AM02, bar length $L_{phase}$)
uses the radial 
profile of the phase of the $m=2$ component which, in the bar region, 
 should be approximately constant. 
Outside it, the phase either shows a
coherent increase or decrease due to a spiral, or large
variations in the absence of clear structures. 
Following AM02, we calculate the phase of the $m=2$ Fourier
component of: (i) the whole bar-disc system and (ii) as a function
of radius, and define 
the bar length  as the maximum radius where these two quantities
differ by less than
arcsin(0.3). In this way, we obtain bar lengths of $8.8$ kpc for Aq-C
and $8.0$ kpc for Aq-G (see Fig.~\ref{fig:phase} and dashed circles in Fig.~\ref{fig:isodensities_z0}).
%as can be inferred from Fig.~\ref{fig:all}.

For the second method 
({\it iv} in AM02, bar length $L_{m=2}$) 
we use the relative $m=2$ Fourier component. If 
the disc and bar were rigid, the length of the
bar would be the radius at which the $m=2$ component
goes to zero. 
This is not the case
in simulated bars, but we can still get an estimate of
the bar length from the radius where the $m=2$ amplitude drops to
a given (arbitrary) fraction of the maximum, that we take to
be $25\%$. 
In this way, we obtain lengths of $8.4$ kpc for Aq-C and of
$7.7$ kpc for Aq-G  
(solid circles in Fig.~\ref{fig:isodensities_z0}).

The third method 
({\it v} in AM02, bar length $L_{prof}$)
uses the density profiles  shown in Fig.~\ref{fig:all}. 
Namely, we take the difference between the profiles
along the major and minor axes, which is zero at the centre,
increases until a maximum and then drops again. If the disc and bar
were rigid, the end of the bar would be where the two projected density
profiles become equal again. In simulated bars, 
the bar length is defined as
the radius where the difference between the profiles
drops to $5$ per cent of the maximum. 
This method yields bar lengths of
 $8.5$ kpc for Aq-C and of  $6.8$ kpc for Aq-G
(dot-dashed circles in Fig.~\ref{fig:isodensities_z0}).

The bar lengths obtained by our three methods agree very
well between them for Aq-C and reasonably well
for Aq-G. For Aq-C we get a mean value of
$\sim8.5$ kpc and for Aq-G of $\sim7.5$ kpc.
%, both well
%within the range of observed bar lengths \citep{Gadotti07, Gadotti11}. 

We can compare our results  to the observations of
$\sim 300$ barred galaxies presented in \citet[hereafter G11]{Gadotti11}.
By applying
bulge-disc-bar decompositions with the BUDDA code \citep{Gadotti08},
G11 obtained bar lengths ($L_{\rm bar}$) and 
ratios between the bar length and the disc scale-length ($L_{\rm bar}/h$).
We have applied the same method to our simulated galaxies (S10), 
obtaining the values of the disc scale
lengths. %, in particular for the two galaxies studied here. 
We find  $L_{\rm bar}/h$ values of
$0.80 - 1.17$ for Aq-C, and of $0.89 - 1.15$ for Aq-G (depending
on the bar length estimation). 
These are at the {\it low} end of the distribution given in G11, but
still compatible with it. On the other hand, if we make
comparisons directly with the
values of bar length in kpc, we find that they are at the {\it high} end of the distribution
in G11 but, again, still compatible with it.
%On the other hand,
%the bar lengths of both simulated galaxies are consistent with
%the results of G11, although
%they are at the long side of the distribution. 
%So it is in the
%high end of one type of comparison and in the low end of the other

%\begin{figure*}
%{\includegraphics[width=35mm]{figures/isovel_stars_x_1_128.eps}\includegraphics[width=35mm]{figures/isovel_stars_45_1_128.eps}\includegraphics[width=35mm]{figures/isovel_stars_y_1_128.eps}\includegraphics[width=9mm]{figures/legend_isovel_vert_1_128.eps}}
%{\includegraphics[width=35mm]{figures/isovel_stars_x_0_128.eps}\includegraphics[width=35mm]{figures/isovel_stars_45_0_128.eps}\includegraphics[width=35mm]{figures/isovel_stars_y_0_128.eps}\includegraphics[width=9mm]{figures/legend_isovel_vert_0_128.eps}}%

%\caption{Velocity fields of the stellar component for Aq-C and Aq-G,
%for three orientations of the bar: along the ordinate (left-hand panel), at 45
% degrees to it (middle panel) and along the abscissa (right-hand panels). 
% The line of sight is always along the ordinate. The colors
%represent the mean velocities along the vertical axis and the white lines are the corresponding isodensity contours.
%}
%\label{fig:faceonv}
%\end{figure*}

\begin{figure}
{\includegraphics[width=32mm]{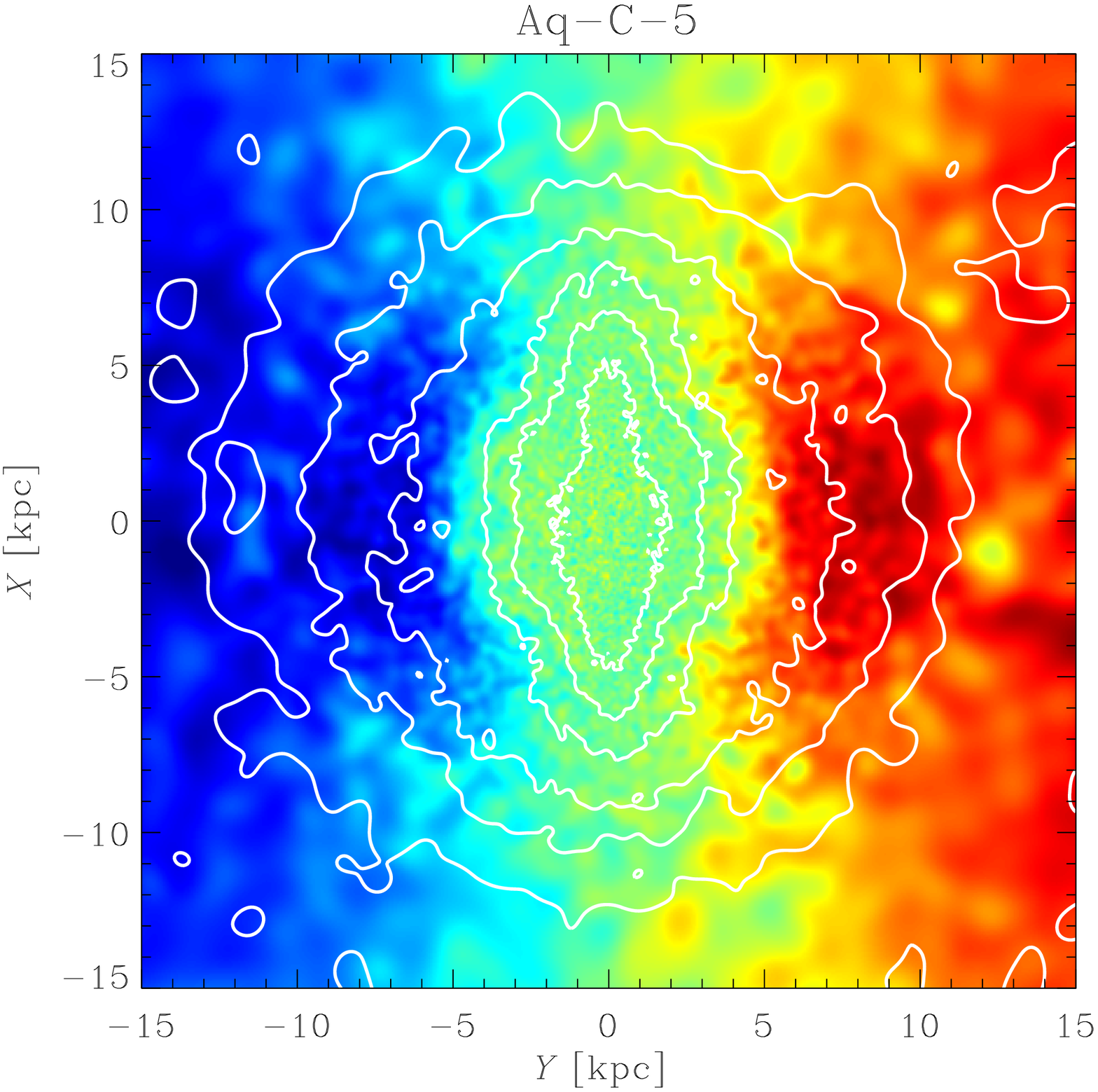}\includegraphics[width=32mm]{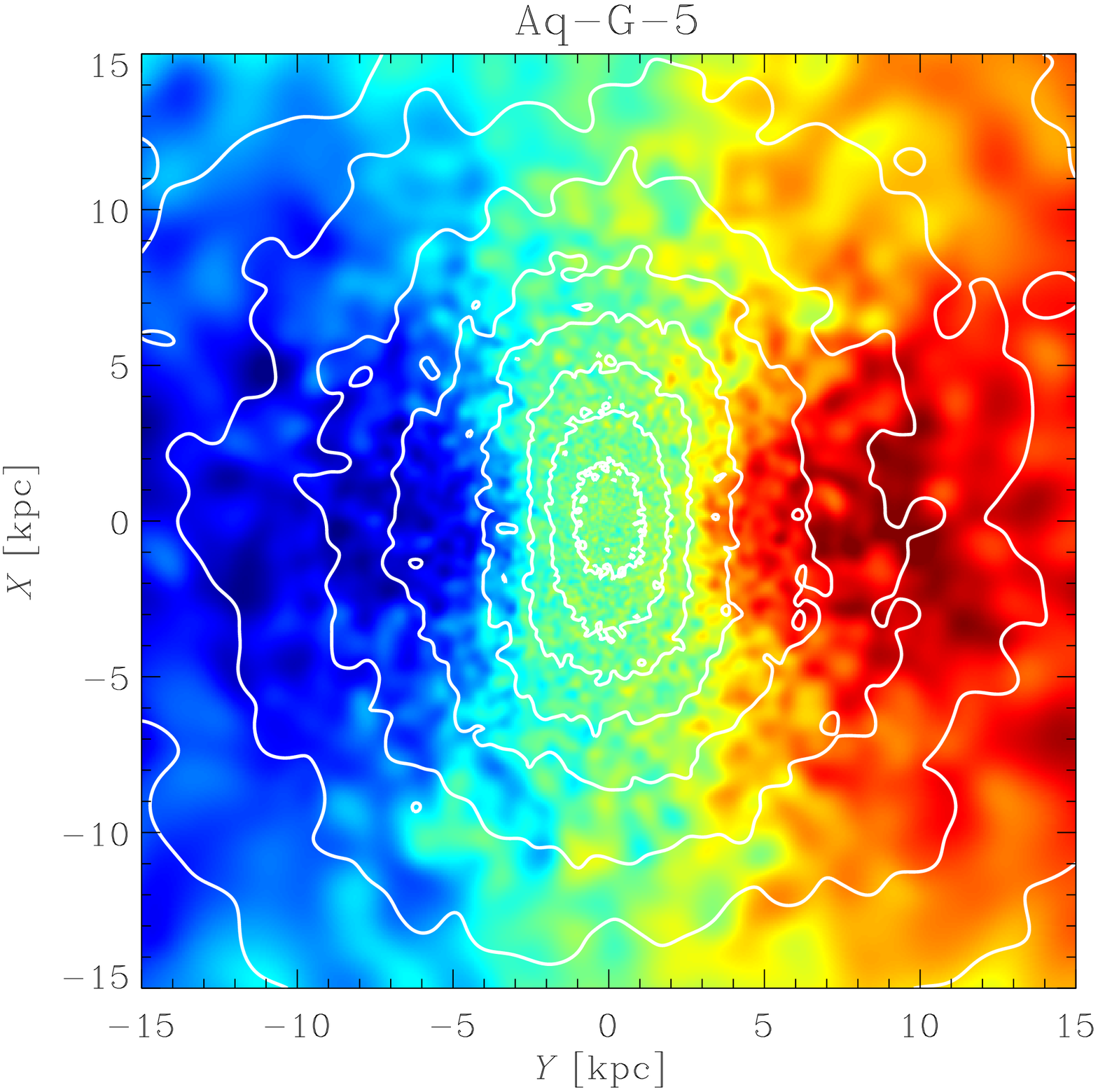}}

{\includegraphics[width=32mm]{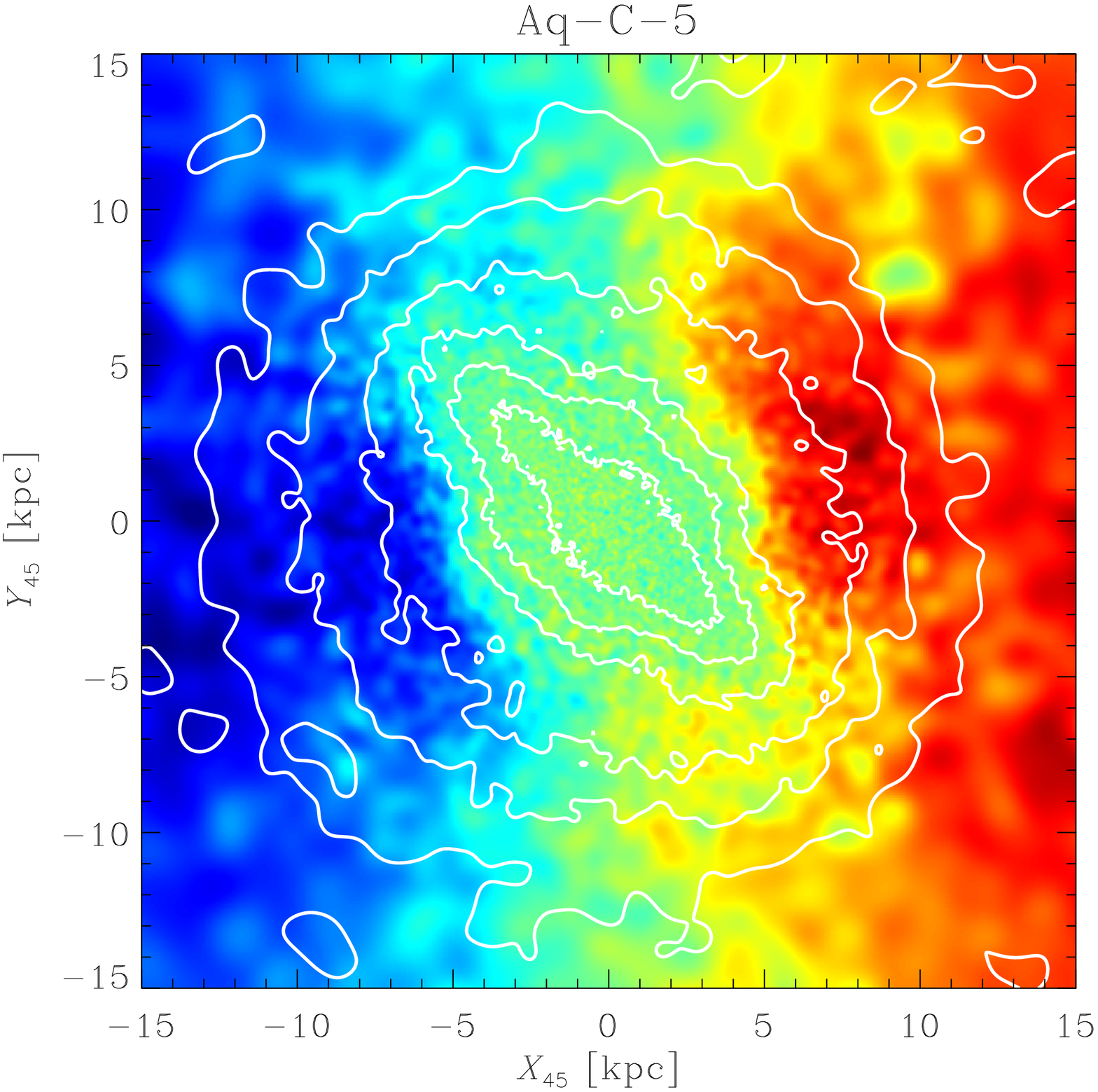}\includegraphics[width=32mm]{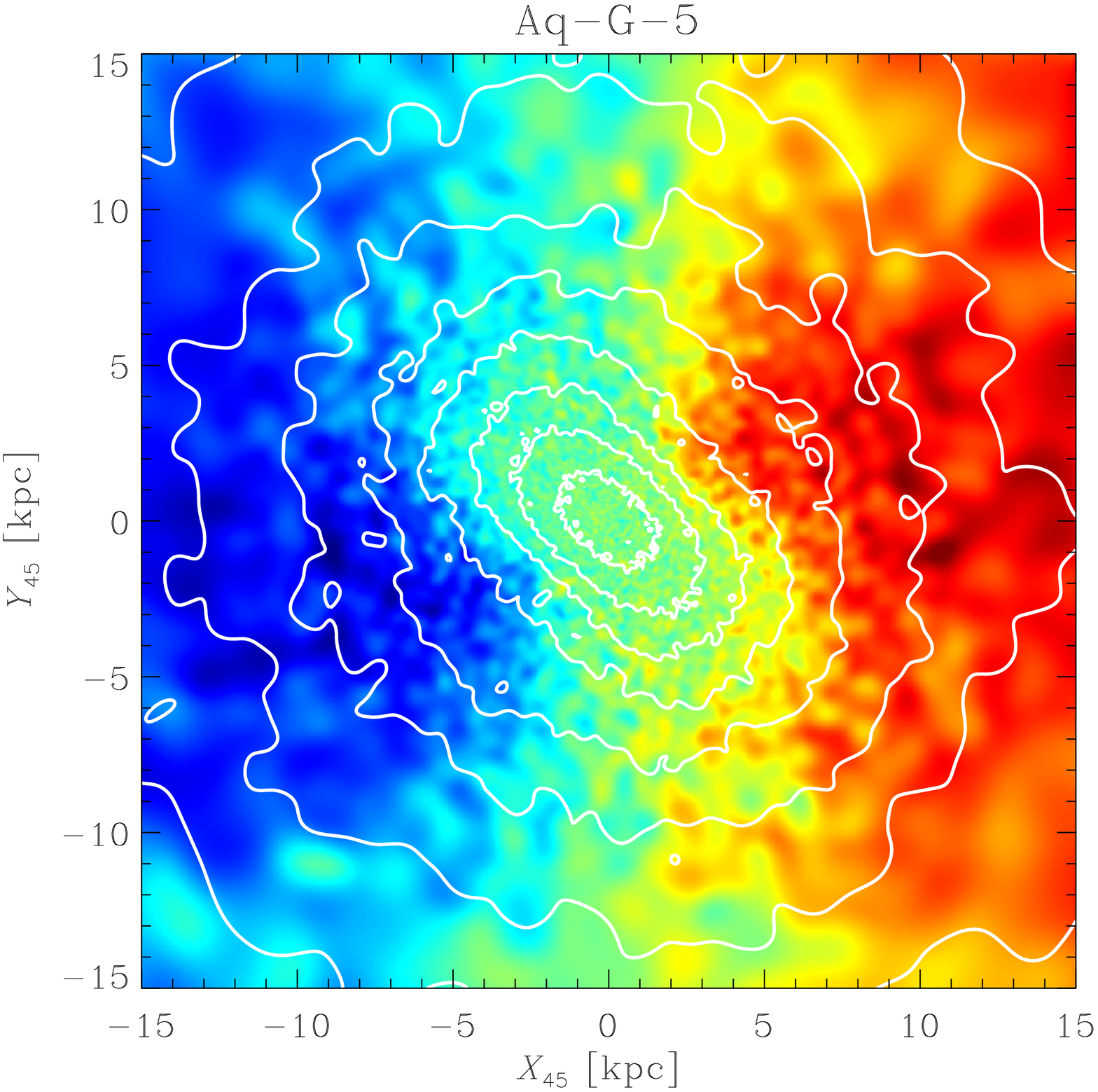}\includegraphics[width=8mm]{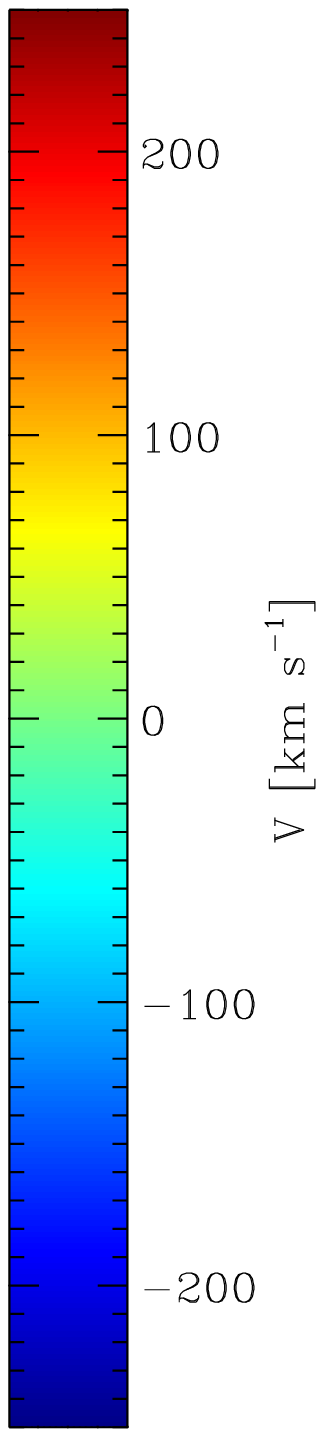}}

{\includegraphics[width=32mm]{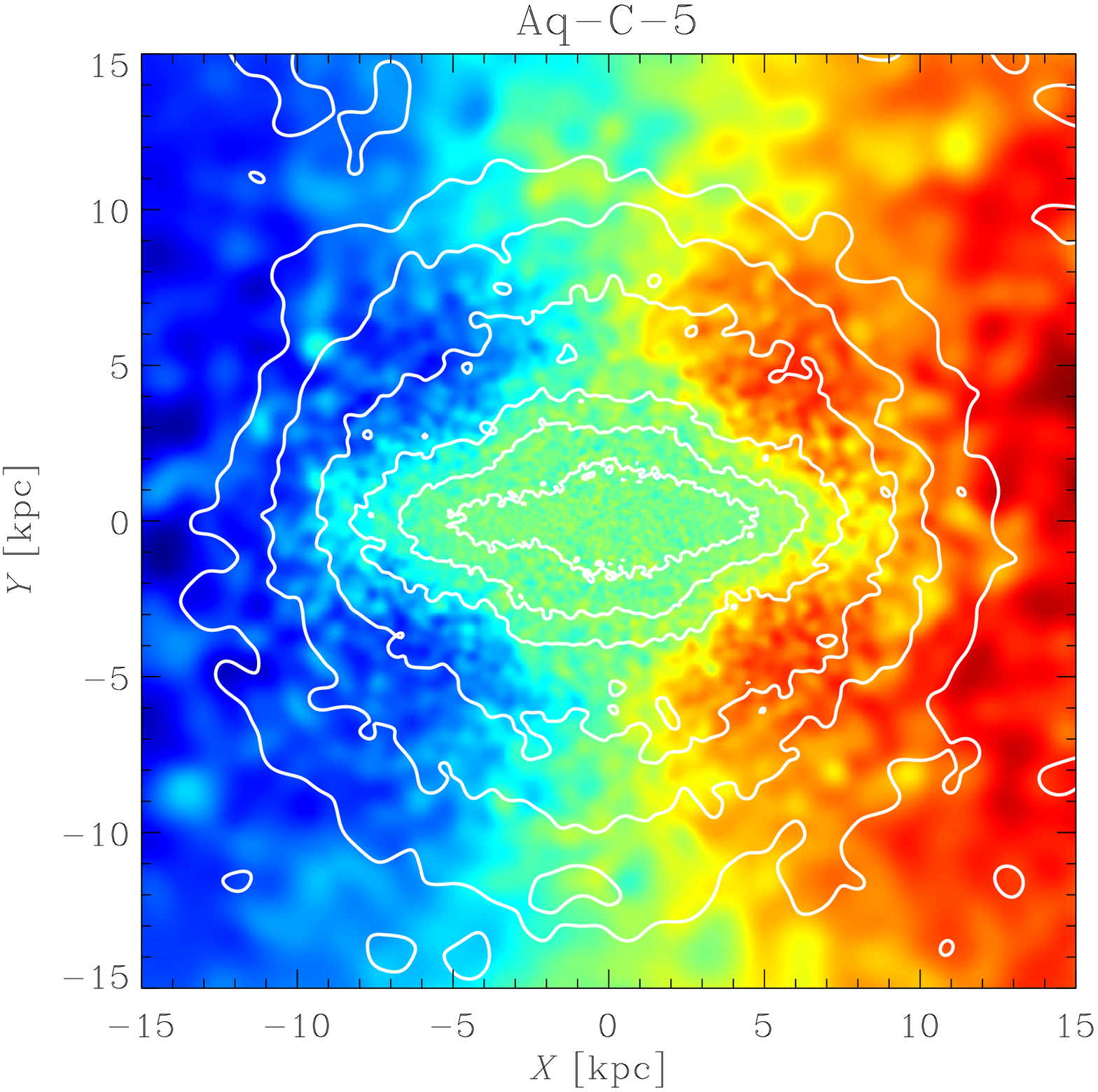}\includegraphics[width=32mm]{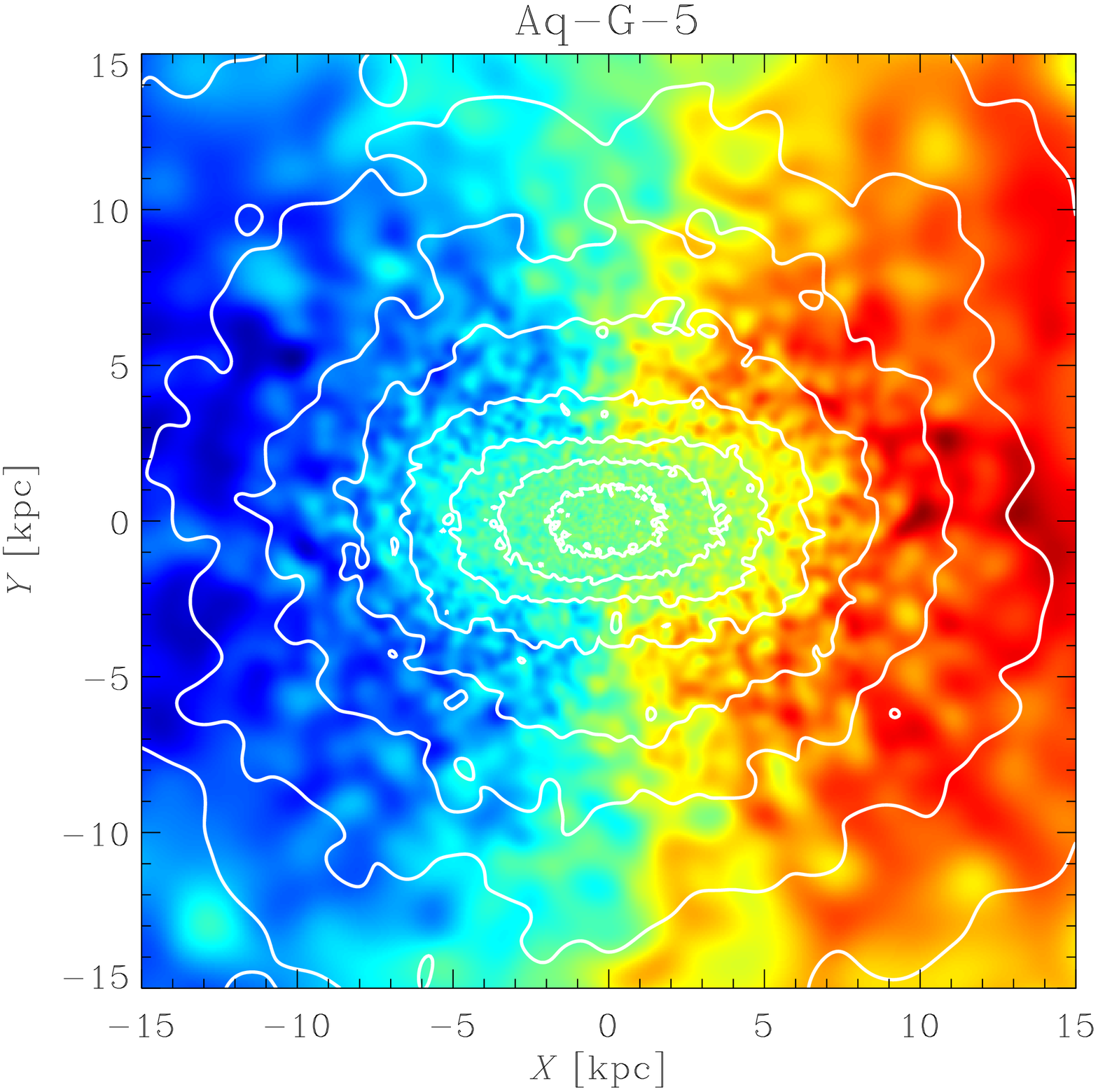}}

\caption{Velocity fields of the stellar component for Aq-C and Aq-G,
for three orientations of the bar: along the ordinate (upper panels), at 45
 degrees to it (middle panels) and along the abscissa (lower panels). 
 The line of sight is always along the ordinate. The colors
represent the mean velocities along the vertical axis and the white lines are the corresponding isodensity contours.
}
\label{fig:faceonv}
\end{figure}

\vspace{-0.25cm}
\subsection{Kinematic properties}

The most telling characteristics of a bar
in a radial velocity field are the presence of kinematical axes
which are not perpendicular to each other, as well as a zigzag (Z)
behaviour of the isovelocity curves corresponding to the systemic velocity
of the galaxy. This Z aspect is best seen in velocity fields where
the bar makes an angle of 45 degrees with the major axis of the
galaxy (e.g. Athanassoula 1984). 
%The orbits of particles in the bar region are elongated along the bar, and
%this reflects itself on the galaxy velocity field. 
To investigate the kinematic structure of our simulated
galaxies, we show in Fig.~\ref{fig:faceonv}  
their stellar velocity fields, obtained as in AM02, i.e. in a way that is
the most convenient for comparing with
observations of galaxies at intermediate inclinations,
% for which the
%contribution of the $Z$ component of the velocity is relatively small
and also allows direct comparison with the AM02 results.
For Fig.~\ref{fig:faceonv}, we rotate the galaxy according to the desired viewing
angles, project all particles on the equatorial plane, 
and observe their line-of-sight (i.e. along the corresponding
plotted ordinate axis) 
velocity component.
% and plot the
%corresponding isovelocities. 
For the three
views shown in the figure, this is equivalent to observing along the bar major
axis, at 45 degrees to it and along the bar minor axis, respectively.
Note that Fig.~\ref{fig:faceonv}
shows the {\it stellar} velocity fields, which
 are not directly comparable to
gaseous ones.

%The velocity fields are analogous to those given by stellar observations and 
%by previously published models. 

The zigzag (Z) behaviour mentioned above is clearly seen in our two simulated
galaxies. The effect in Aq-C is quite strong, and is comparable to that
in model MH of AM02. The effect is weaker in Aq-G, but
still stronger than the corresponding MD model of AM02.
When we observe along the bar major axis,
the isovelocities show the standard spider diagram form. The central
concentration is less than in the dynamical simulations of AM02, but
this is due to our larger softening. 
%The intermediate angle velocity
%field shows the `$Z$' structure characteristic of barred galaxy velocity
%fields (see e.g. \citealt{Peterson1978}, albeit for gas). 
Finally, when we
view along the bar minor axis, the velocity fields have an
inner region which roughly delineates the bar and where the velocity
is lower. This was also found in AM02 and is due to the
fact that near the ends of the bar the particles are at their
apocenters, i.e. the velocities are lower.
The mean velocities in those regions can be further lowered
if the corresponding periodic orbits have loops at their apocenters 
\citep{Athanassoula1992}.

\section{Conclusions}\label{sec:conclusions}

We presented the first study of bars  formed in fully
cosmological, hydrodynamical simulations of Milky Way-sized
haloes. % in a $\Lambda$CDM universe. 
The simulations 
%follow the formation of two galaxies of
%similar mass to the Milky Way, including 
include star formation,
metal-dependent cooling, feedback from supernova and a
UV background field. 
In particular, we investigated the morphology, strength
and length of the bars,
their projected density profiles
%along their major and minor axes, and the azimuthally-averaged 
%density profiles. 
and kinematical properties.

The strongest of our two bars formed in a bulge-disc-halo
system (Aq-C), while the weakest  in a galaxy with a disc but no significant
bulge (Aq-G). Compared to dynamical simulations, the strongest bar
is similar to those found in systems where a considerable amount
of angular momentum is exchanged between the bar and the halo.
In contrast,
the weakest bar is more reminiscent of dynamical simulations where
considerably less angular momentum has been redistributed.
The bar strength difference between our two simulated
bars could be due to the effect of the
bulge on the angular momentum exchange (which is negligible in
Aq-G), as found in pure N-body
dynamical simulations of isolated discs (AM02, A03), but the effect of
interactions and/or of the gas component could also be decisive.
%We analysed the simulated bars at $z=0$, both qualitative
%and quantitatively,  applying methods already 
%in use for bars in dynamical
%simulations and in observations, so that a direct comparison with
%previous results was possible. 

The Aq-C (strongest) bar is very thin and rectangular-like,
while that of Aq-G (weakest bar) is very fat, and also
rectangular-like. Rectangular shapes
are observed in strongly barred galaxies, so Aq-G is, in this respect,
like an hybrid between strong and weak bars.
Another difference is that
the bulge of Aq-C has a rather peculiar shape which
has so far not been observed. The peculiar shape 
is presumably a result of the cosmological formation of our
galaxies, where mergers and/or interactions are common at
all times.

The lengths of our bars are in relatively good agreement with observations
of G11, although
in the long tail of the distribution. The strongest
bar, formed in our most massive galaxy, is also the longest.
On the other hand, the ratios between bar length and disc scale-lenght
are at the low end of the observed distribution, but still compatible
with it.

The density profiles are similar to those found in dynamical simulations;
however, we detect some differences particularly in Aq-C
and in the very central regions, probably because
we include stars from all components in our analysis,
unlike 
in dynamical simulations.
%Higher resolution cosmological simulations, available in the
%near future, 
%will allow further and more detailed comparisons.   
Finally, the kinematic properties of our bars are similar
to those observed and to those found in dynamic simulations.

The fact that the bars were obtained in a cosmological
setting, and that their properties agree relatively well with
known properties of  bars,  is not minor. Our galaxies
grow significantly from $z=3$ to $z=1$, where the bars
are first detected in the simulations. Moreover, external
(accretion, interactions and mergers) and internal
(cooling, star formation and feedback) effects strongly affect the
galaxies  during their evolution.
%Of course, if we look closely enough, differences become clear,
%the most flagrant being the shape of the isodensities of the
%bulge in the face-on view of simulation Aq-C. These do not look
%sufficiently circular-like and will influence somewhat the 
%Fourier amplitudes in the inner parts. 
%Despite the differences described above between our simulated
%bars and real galaxies and dynamic simulations,
For these reasons, our results are quite
satisfactory and very encouraging, and should incite
more work on the study of bar formation in a cosmological context.

\vspace{-0.3cm}

\bibliographystyle{mn2emod}

\bibliography{biblio.bib}

\end{document}